\documentclass[fleqn]{book}
\usepackage{yearbook,natbib,latexsym,epsfig,graphics}

\def\eps@scaling{.95}
\def\epsscale#1{\gdef\eps@scaling{#1}}
\def\plotone#1{\centering \leavevmode
    \epsfxsize=\eps@scaling\columnwidth \epsfbox{#1}}
\def\plotfiddle#1#2#3#4#5#6#7{\centering \leavevmode
\vbox to#2{\rule{0pt}{#2}}
\includegraphics{#1}}

\def\kms{\ifmmode {\rm\,km\,s^{-1}}\else
    ${\rm\,km\,s^{-1}}$\fi}
\def\ms{\ifmmode {\rm\,m\,s^{-1}}\else
    ${\rm\,m\,s^{-1}}$\fi}
\def\kmsMpc{\ifmmode {\rm\,km\,s^{-1}\,Mpc^{-1}}\else
    ${\rm\,km\,s^{-1}\,Mpc^{-1}}$\fi}
\def\hkmsMpc{\ifmmode {\rm\,h^{-1}\,km\,s^{-1}\,Mpc^{-1}}\else
    ${\rm\,h^{-1}\,km\,s^{-1}\,Mpc^{-1}}$\fi}
\def\lya{{\rm Ly}$\alpha$}

\def\msun{\ifmmode {\rm\,M_\odot}\else ${\rm\,M_\odot}$\fi}
\def\Msun{\ifmmode {\rm\,M_\odot}\else ${\rm\,M_\odot}$\fi}
\def\lsun{\ifmmode {\rm\,L_\odot}\else ${\rm\,L_\odot}$\fi}
\def\Lsun{\ifmmode {\rm\,L_\odot}\else ${\rm\,L_\odot}$\fi}
\def\rsun{\ifmmode {\rm\,R_\odot}\else ${\rm\,R_\odot}$\fi}
\def\Rsun{\ifmmode {\rm\,R_\odot}\else ${\rm\,R_\odot}$\fi}

\def\cmtw{\ifmmode {\rm\,cm^{-2}}\else ${\rm\,cm^{-2}}$\fi}
\def\cmthr{\ifmmode {\rm\,cm^{-3}}\else ${\rm\,cm^{-3}}$\fi}

\def\ergps{\ifmmode {\rm\,erg\,s^{-1}}\else ${\rm\,erg\,s^{-1}}$\fi}
\def\ergpscmtw{\ifmmode {\rm\,erg\,s^{-1}\,cm^{-2}}}

\def\cf{{\it cf.~}\ }
\def\eg{{\it e.g.}}

\def\deg{\ifmmode {^{\circ}}\else {$^\circ$}\fi}
\def\degr{\ifmmode {^{\circ}}\else {$^\circ$}\fi}
\def\degs{\ifmmode {^{\circ}}\else {$^\circ$}\fi}

\def\etal{{\it et al.~}}

\def\Ho{\ifmmode {\rm\,H_\circ}\else ${\rm\,H_\circ}$\fi}
\def\hnot{\ifmmode {\rm\,H_\circ}\else ${\rm\,H_\circ}$\fi}
\def\h0{\ifmmode {\rm\,H_\circ}\else ${\rm\,H_\circ}$\fi}
\def\hnotunit{\ifmmode {\rm\,km\,s^{-1}\,Mpc^{-1}}\else
    ${\rm\,km\,s^{-1}\,Mpc^{-1}}$\fi}
\def\qnot{\ifmmode {\rm\,q_\circ}\else ${\rm q_\circ}$\fi}
\def\q0{\ifmmode {\rm\,q_\circ}\else ${\rm q_\circ}$\fi}

\def\arcsec{\ifmmode {^{\prime\prime}}\else $^{\prime\prime}$\fi}
\def\asec{\ifmmode {^{\prime\prime}}\else $^{\prime\prime}$\fi}
\def\arcmin{\ifmmode {^{\prime}}\else $^{\prime}$\fi}
\def\amin{\ifmmode {^{\prime}}\else $^{\prime}$\fi}

\def\hetwo{He {\small II}}

\def\otwo{O {\small II}}
\def\civ{C {\small IV}}
\def\h{{\rm h}}
\def\lesssim{\mathrel{\hbox{\rlap{\hbox{\lower4pt\hbox{$\sim$}}}\hbox{$<$}}}}
\def\gtrsim{\mathrel{\hbox{\rlap{\hbox{\lower4pt\hbox{$\sim$}}}\hbox{$>$}}}}
\let\ga=\gtrsim

\begin{document}

\Chapter{The Most Distant Galaxies}

\author{Hyron Spinrad}

\address{Department of Astronomy University of California, Berkeley,
California, U.S.A. 94720-3411\hfill\break\url{e-mail:
hspinrad@astro.berkeley.edu}}

\begin{abstract}

I review selected current observations of distant galaxies and our
interpretation of the fragile (and occasionally contradictory)
data. Galaxies at the ``contemporary limit'' of technology and
redshift ($z \sim 6$) are difficult to locate in the first
place. Moreover, the large redshift may push some critical confirming
and/or interpretative analysis toward unfamiliar IR wavelengths. I
will concentrate on observational means and results to explore the
early evolution of galaxies. We also note the biases that intrude on
plans for the interpretative aspects of distant galaxy photometry and
spectroscopy. We discuss the best methods of selection for those very
distant systems; these methods include utilizing strong sub-mm
emission from dust, photometry indicating a UV "spectral break," and
finally the signal of a strong \lya\ emission line. This feature has
now carried us to a galaxy redshift in excess of $z = 6.57$!
\end{abstract}


\section{Introduction, Motivations, and Questions} \label{intro}

The study of distant galaxies is empirically demanding -- not
surprisingly, as these galaxies are very faint.

Of course there are a variety of motivations to observe and perhaps
understand distant ``units of the Universe''. We would like to detail
the present-day ``lumpiness'' of the Cosmos and its evolution from a
very smooth ``sea'' at decoupling. At the nominal redshift of the Cosmic
Microwave Background the key fluctuations on reasonable scales are
only of order $10^{-5}$. Of course at $z \sim 0$ we have a very
inhomogenous distribution of baryons we call galaxies and the
Intergalactic Medium (IGM hereafter).

Noting the obvious, studying distant galaxies is synonymous with
traveling far back in cosmic time towards the birth of massive
sub-structures and large galaxies. Can we now see directly the
development of single galaxies of Milky Way dimensions?

We now believe that most galaxies form and accumulate either (1) by
the infall of gas (and dark matter) as ``monolithic''
entities, self-gravitating by the time we can observe them, or (2) by
a series of major and/or minor mergers. This is the now-popular
``bottom-up'' scenario. Here it is presumably difficult to
catch the small and immature systems in the act of merging,
depending perhaps on the appropriate dynamical time scales. Thus for
scenario (2) we would anticipate young galaxies to illustrate complex
morphologies, quite different from those of the mature galactic
systems we study readily here and now, at zero redshift. There is
indeed some evidence for ``recent'' mergers from the fine images of
distant galaxies observed with the Hubble Space Telescope (HST) - see
Stern \& Spinrad (1999) for some plausible early merger examples (Fig. 1). And
we'd like to push these examples back in cosmic time to even
``younger'' galaxy growth - but the first problem is, quite naturally,
the location of small and dismally faint candidates for galaxies in
formation

\begin{figure}[h!] 
\centering
\epsscale{0.9} 
\vspace{-0.25in}
\plotone{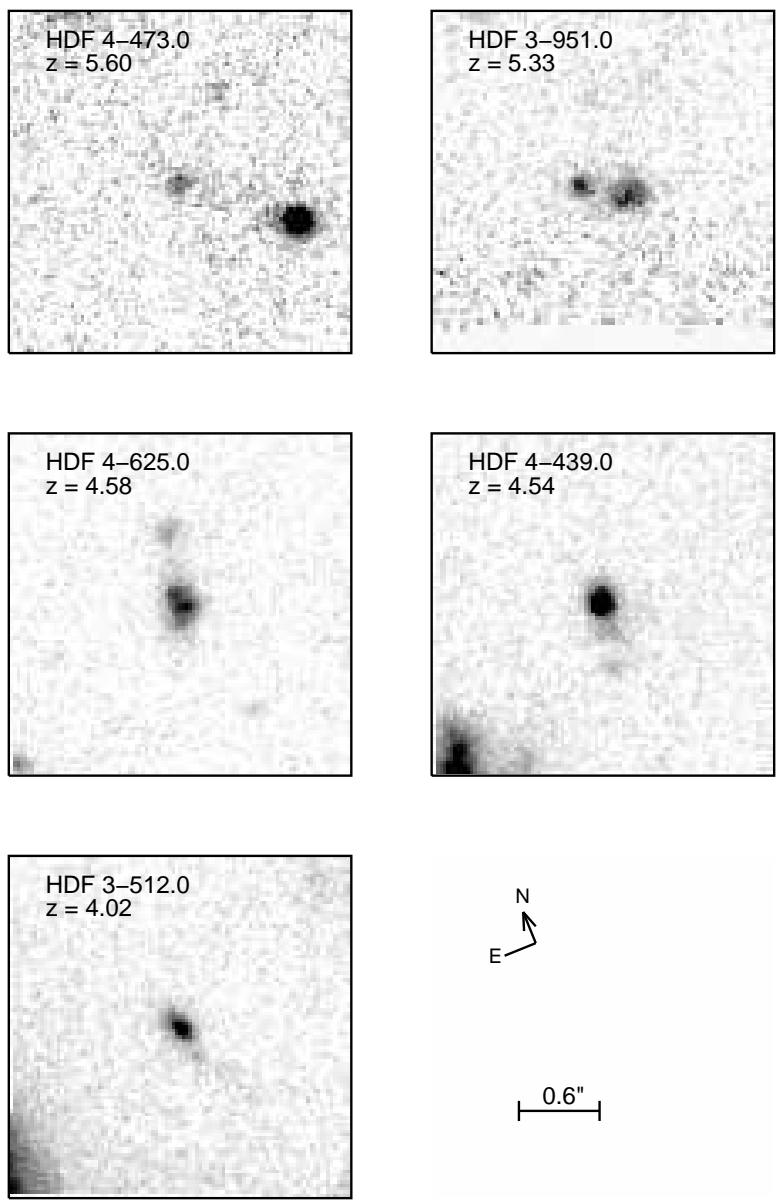}
%\plotone{ewds_sl_int_tidez_lett2.eps}
\vspace{-0.1in}
%\plotfiddle{3b_vs_N.ps}{3.6in}{0}{50}{50}{-150}{-73} % 
\caption{HST images of five spectroscopically confirmed galaxies
located in the HDF(N). Note the distortions, small tails, and multiple
centeral components - presumably due to mergers. Overall the galaxies
are obviously quite small at this stage of their evolution.  From
Stern \& Spinrad (1999)}
\end{figure}

Another important contemporary research area emerging is the study of
intergalactic (gaseous) matter usually seen in silhouette
against a bright background source like a QSO or an unusually bright
and distant galaxy . And now, new observational techniques are
beginning to tell us about the interaction history of galaxies and the IGM
(\emph{cf.} Adelberger \etal, 2003).

One of this paper's topics, directly or indirectly stated, is just how
early in cosmic epoch (parameterized by redshift) we can study
individual galaxies or their ``pre-galactic'' fragments. There is only a
short time interval between the early epochs beyond $z = 3$ (see Figure
2). How can the galaxies evolve so quickly?

\begin{figure}[h!] 
\centering
\epsscale{0.9} 
\vspace{-0.25in}
\plotone{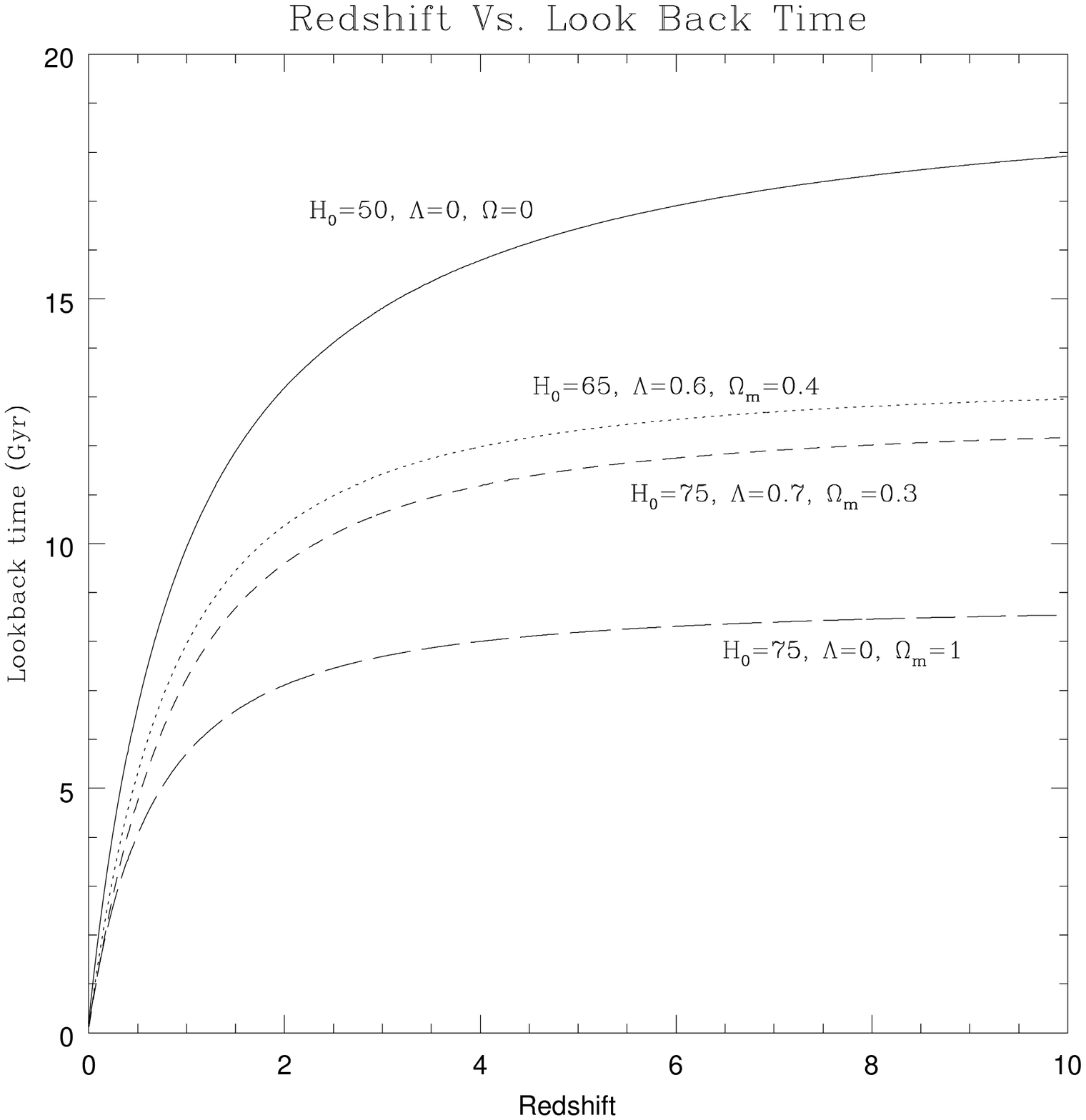} %! \sim/vofz/...
\vspace{-0.1in}
%\plotfiddle{3b_vs_N.ps}{3.6in}{0}{50}{50}{-150}{-73} % 
\caption{A plot of look back time (in Gyrs) versus redshift for three
cosmological models. Most might now prefer the short-dashed curve.
Note that at high $z$ ($z \ga 3$) the time intervals become quite
short.  Figure by Curtis Manning}
\end{figure}

The historical view of our empirical and theoretical march outward
toward higher redshift has shown a fairly rapid expansion. By 1976 a
few radio galaxies had been located and studied at $z > 0.5$. The $z =
1.0$ threshold (for galaxies) was crossed in 1981. Of course Quasars
and QSOs had been actively observed and known earlier at large
distances - redshifts in the 1960s and 1970s taking us to $z = 2.01$
(Schmidt 1965) and then 2.88, and then to $z = 3.5$ (OQ 172;
Baldwin \etal, 1974). Finally, $z = 4$ for QSOs was surpassed by the
Palomar two-color-based searches (Schneider, Schmidt \& Gunn 1991),
and searches for \lya\ on low-resolution grism spectra
(Osmer 1999) were equally successful. Almost all the recent
stages of the ``QSO-z race'' have emphasized red-IR photometry and
unusual colors, since the $z \sim 5$ QSOs are heavily depressed by the
\lya\ forest of the IGM (see Fan et al. 2001). The largest published
QSO redshift to date is $z = 6.28$ (Fan \etal, 2002; Pentericci \etal,
2002a).

Now we are witness to the era of a friendly race toward higher and
record-breaking galaxy redshifts. The current limit for galaxies,
which we shall detail later in this publication, is near $z = 6.5$. Is
this redshift close to the end of the ``dark ages'', where re-ionization
by massive stars and/or early QSOs play as vital sources of ionizing
radiation? We return to this topic, with empirical evidence, toward
the conclusion of this review.

\section{Some Issues In The Contemporary Theory Of Early Galaxy Evolution}

Over the last three or four years, the thoughts of theorists have
narrowed on the birth and evolution of galaxies - including dark
matter halos, plus the baryons we observe more directly. These
adventuresome researchers have bi-modally attacked the problems with a
pair of model types. Most contemporary modeling assumes, \emph{ab
initio}, the Lambda Cold Dark Matter cosmology (LCDM).

Following Weinberg \etal, (1999), we note that the current (broad) theory
of galaxy formation and early evolution follows White \& Rees (1978) and
their ``successors'' - gravitational collapse of a dark matter halo,
gas falling into the potential well so defined, and then gas
astrophysics (cooling, contracting, and eventually forming stars in a
dense baryonic core). Now we often add inflationary cosmological
parameters and thus demand $\Omega_m + \Omega_{\Lambda} = 1$.

The ``technology'' for modeling often takes one of two paths. The first
is hierarchical numerical simulations (with a realistic treatment of
the collapse) including additional gas-phase physics and plenty of
computational effort to cover the wide size range of non-spherical
assemblies (the ``roots'' of the assembly ``tree'') that appear.

The second tool, deemed the semi-analytical approach, assumes again
LCDM halos. The proto-galaxies contract within, and then we find small
sub-galactic systems (or fragments?) with the specific
physically-motivated ``stories'' given by the strengths of their
star-burst mergers.  The mergers obviously increase the model masses,
and also modify the relative numbers of luminous stars and the amount
of residual gas.  Even before that step, the semi-analytic models
utilize the Press-Schecter (Press \& Schechter 1974) formalism to describe the
number of halos as a function of their mass. This approach
allows conventional and mature use of population synthesis and even
chemical evolution schemes in conjunction with the mergers demanded to
build up galaxies of reasonable mass with moderate star-formation
rates.

One of the strengths of the direct numerical simulations is to utilize
the non-spherical distribution of dark matter and baryons to produce a
more realistic treatment of the model's gravitation. Then the more
``astrophysical'' computations can proceed; Weinberg \etal, (1999)
predict the surface densities of galaxies as a function of
their star-formation rate (SFR) over a relevant range of redshifts.

The semi-analytic models (\emph{cf.} Baugh \etal 1999; Somerville \&
Primak 1999) have now been amplified to include a range of interesting
physical processes, hopefully relevant to early galaxy evolution. For
example, the central baryons and the outer dark matter (DM) halo
interact to change the halo structure and foster further contraction
of the model galaxy.  Baugh \etal (1998) mention that the main
constraining property of local galaxies they favor for comparison with
semi-analytic modeling is the field galaxy luminosity function. The
agreement is good; one can then easily visualize the effect of
omitting or including various individual physical processes, like
star-formation (SF) feedback.

The SFR in the early Universe (say, to $z = 3$ or 4) of these models
is also well-matched by observations of the SFR per unit
volume. (Madau et al. 1999).

Weinberg et al. (1999) also show the numerical simulation's cumulative
distribution of galaxies (with the parameter $=$ surface
density/$\Box^{\prime}$/unit $z$) as a function of their SF rate from
$z = 10$ to $z = 0.5$.  At the moderately large galaxian ${\rm SFR} =
10 ~\Msun\,{\rm yr}^{-1}$ for $z = 5$, the predicted surface density
of galaxies is nearly $5/{\Box^{\prime}}$.  This surface density is
rather higher (by a factor of $\sim 3$) than observed by Spinrad and
collaborators (although some of this observational statistic is
derived from the \lya\ - SFR correlation, which may be suspect).  The
best unpublished observational estimate for the SFR surface density at
$z \sim 5$ is now $2 \pm 1/{\Box^{\prime}}$.  However, this surface
density for \lya\ emitters is uncertain because their continua are
often very weak and thus not necessarily sampled consistently in terms
of galaxy luminosity.  The theoretical simulations and follow-up
astrophysical scaling may, of course, be systematically over-efficient
in, for example, converting cooling gas to massive star births.

The numerical simulations with LCDM may have one flaw: they
over-predict the number of small galaxies near larger ones (which are
countable) and thus the number of stars at low redshifts. We are not
positive that a real problem exists; it may be that dark halos with
coupled non-stellar baryons (\eg, high velocity clouds
(Klypin \etal, 1999) are being ``counted'' as observable systems.

The potential problems of early galaxy evolution from the theoretical
side may well change, increasing or decreasing as their confrontations
with empirical ``facts'' or new understandings go forward. The general
outlines of the theory and relevant observations are probably fairly
firm.

\section{A Race For The Maximum Redshift}

It is a very human tendency to climb a celestial mountain. So it
stands for any race, including that of finding individual objects at
greater and greater distances, abbreviated usually as at larger
redshift, or ``bigger z'' (where $1+z = \lambda_{\rm
observed}/\lambda_{\rm emitted}$).

As Stern \& Spinrad (1999) pointed out in their Table 1, there has been a
fairly rapid increase in ``zmax'' for galaxies; we went from $z =
0.20$ in 1956 to $z = $1 by 1982, but then to $z = 5.3$ in 1998 and $z
= 5.7$ in 1999. The record-breaking progress since 1998 has been due
to observations of the strong \lya\ (from rest $\lambda 1216$ \AA)
emission line, shifted to the visible and red by the Universal
expansion. Over the past year the ``LALA'' (Large Area Lyman Alpha
survey) team (Rhoads \etal 2003) have selected \lya\ emitters to $z =
5.75$.  They are currently taking images for the $z=6.6$ airglow window.
Also in 2002 Hu et al. have located a cluster-lensed galaxy at the
outstanding redshift of $z = 6.56$! And as these pages are completed,
a Subaru group has found a faint \lya\ galaxy at $z = 6.578$.

Modern research on quasars (QSOs, to be more precise), has also
progressed; Osmer (1999) reviewed the situation 3 years ago, with QSOs
located up to $z = 5.0$. Since then, the Sloan Digital Sky Survey has
been successfully pushed QSO redshifts to and beyond $z = 6$! The key
here is to obtain good red and near IR photometry, in particular
looking for objects with very red (I-z) colors. The Sloan results are
very current; Fan \etal\ (2002) found SDSS J103027.1 at $z = 6.28$, and
a preprint on another Sloan QSO at $z \sim 6.43$ is just available as this
section is being written. So the most distant QSO to date still trails
the most distant, much fainter normal galaxy by a modest margin!

The distant QSOs are likely buried in a host galaxy which itself is
well-hidden in the glare of the Active Galactic Nucleus (AGN). We now
assume the presence of the underlying galaxy of stars and gas, in part
confirmed indirectly by the normal abundances of the elements inferred
from the emission lines in the QSO spectra.

\section{The Identification Of Very Distant Galaxies}

How do we go about locating the faint and distant galaxies at the
heart of our exploration and this review?

We found several successful (or partly successful) methods to locate
the faint targets at high redshift: none are without ``flaws''. For
example, some methods are weakened by ``contaminants'', be they
intrinsically faint M, L, or T dwarf stars in the galactic disk, or a
mis-identified (longer wavelength, smaller redshift) emission line.

Following the theme in the Stern \& Spinrad (1999) review, we shall
discuss several of the more successful search techniques; initially
we'll review the finding of distant galaxies utilizing non-optical
wavelengths. Often these techniques turn out to be ``safe'' and
productive.

\subsection{Radio-Loud Galaxies}

Radio galaxies at high redshift are rare but interesting guides to the
location of large, mature galaxies and correlated structures -
sometimes actual (rich) clusters (van Breugel \etal\ 1999; Lilly \&
Longair 1984).  For some specific cases, like 4C 41.17 ($z = 3.798$),
and also radio sources resembling it, we note that steep radio
spectral indices and moderate flux densities correlate with high
redshift and great luminosity.  Such objects are visible across much
of the presently observable Universe.

The stronger radio galaxies, those with fluxes $S_{408}\ge 100$ mJy,
tend to follow a good Hubble relationship in the observer's near-IR
bands; that is, their (K,z) magnitude--redshift correlation is linear
with only a moderate scatter.

This result shows that the powerful radio galaxies, E systems in
morphological appearance, have a fairly strong resemblance to a
luminous ``standard candle'' (van Breugel \etal\ 1999; Best \etal\
1999). The history of the steep radio spectral index ``angle'' is
reviewed by de Breuck \etal\ (2000). Going for the steeep radio
spectral counterparts also tends to minimize the ``contamination'' by
Quasars (radio spectral indices $ < -1.3$).

We then may inquire: are all steep radio sources luminous galaxies and
Quasar candidates? The answer here is mainly negative; it is the
medium strength (so as not to exceed some limiting intrinsic
luminosity) steep spectrum sources, identified at long wavelengths in
the optical and IR that have the greatest promise in pointing out very
distant spectrographic targets. These may be radio-loud stellar
systems at a large redshift, say $z \ge 4$.

Somewhat tangential to our central motivation, we note that at both
small and large distances, radio galaxies possess some/many of the
characteristics of giant E galaxies (or luminous cluster Es). Since
these E galaxies here and now have a strong correlation amplitude at
small separations, we can anticipate many of the distant radio Es to
also have smaller companions - perhaps in a group population. These
earmarks of early structure are going to be valuable; the recent paper
of Venemans \etal\ (2002) illustrates a large (2Mpc) overdense region
at a redshift $z = 4.1$ located ``around'' the radio galaxy
TNJ1338-1942.  So the radio galaxy becomes a valuable marker in such a
case. We note another, less well-documented case in the HDF(N) is
currently being explored by Stern, Dey, Dawson, and Spinrad.  Here the
redshift is even greater; the first observed galaxies have $z \simeq
5.2$. No radio source takes part in that overdensity region,
however. Stern et al. (2003) show a group surrounding the radio galaxy
MG0442+0202 at $z = 1.11$.

The record redshift for a radio galaxy is still $z = 5.19$
(van Breugel \etal\ 1999), with TNJ0924-2201. Several observing groups are
concentrating on the identification of deep samples showing a
steep spectrum, with the expectation that some are ultra-luminous and
located at $z > 5$. These are rare systems; one problem in
interpretation is that it should be a fairly slow process to ``build''
a large and luminous galaxy. Perhaps it requires a cosmic interval in
excess of a billion years to do so, either in the model described as a
``monolithic collapse'' (Eggen, Lynden-Bell \& Sandage 1962), or by
the accumulation of smaller structures (Searle \& Zinn 1978) - a
hierarchical model. With the currently popular cosmology [$H_0 = 65$,
$\Omega_{\lambda} = 0.7$, $\Omega_m = 0.3$], the look-back interval
between $z = 4$ and (an arbitrary) $z = 20$ is only $\sim 1.2$ Gyr
(see Fig. 2 again). That might be sufficient time to build a large
galaxy; the implication is then a SFR of $\sim 80 ~\Msun {\rm
yr}^{-1}$. That is a rarely observed and atypically high SFR. So it is
a clue that massive radio galaxies are unlikely to be found at $z >
5$. But the near-IR Hubble Diagram of the highest-z radio galaxies
plotted by van Breugel \etal\ (1999) continues to suggest a continuity in
galaxy luminosity which we may still extrapolate to stellar (and
gaseous) mass similarities. 

Under standard CDM-based models of galaxy evolution, we expect the
giant elliptical galaxies, which are the hosts of today's radio
galaxies, to form late (at $z\sim 1$) through a process of merging of
smaller sub-units. Although these models seem to be consistent with
what is known so far about field galaxy evolution (e.g.\ Barger et
al.\ 1999), and indeed with observations of the hosts of the
radio-quiet quasar population (Ridgway 2000), it is clear that radio
galaxies are an exception. They seem to only show significant
evolution at $z>2$, and still appear to be luminous galaxies at $z\sim
3$ and perhaps beyond. One possible solution is that the most massive
galaxies formed first in so-called anti--hierarchical baryonic
collapse.  In this model (Granato et al.\ 2001) the high baryon
densities in the centers of the most massive dark matter halos cause
them to start forming stars early.  Thus, the fate of simplistic
theoretical analyses suggest the need for a sharper observational
analysis.

\subsection{Galaxies With Strong X-ray Emission (Hidden AGNs)}

To date many new X-ray galaxies have been located, using modern X-ray
satellites such as Chandra and XMM-Newton. However, there are few
X-ray-selected very distant galaxies, or AGN. To my knowledge there is
one at a redshift in excess of 5; it is \#174 in Barger \etal\ (2002) at
$z = 5.186$, in the Chandra Deep-Field, North. We will return to this
galaxy a bit later. There are, however, a considerable number of QSOs
and other clearly noticed AGN at $z > 4$ (cf. Brandt 2002). Why
are we physically interested in X-ray galaxies, anyway? As
Barger \etal\ (2001) affirm, X-ray surveys, especially at hard (2-7keV)
energies, provide a direct indication of an AGN, presumably due to an
ultra-massive black hole at the galaxy nucleus. At $\ga 5$ keV,
absorption will play less of an obscuring role than seen for some
``hidden'' AGNs at optical frequencies and soft X-ray
energies.  Complete samples of hard X-ray energies are now possible
with the Chandra X-ray Observatory; the $1^{\prime \prime}$ X-ray
positions produce robust optical identifications of the
counterparts. And about half of the sources can be identified with
optically bright and ``quiet'' galaxies; they are at small redshifts.

With the longest integrations (say, one mega-second integrations) we
begin to locate the faint X-ray population. Some of these sources are
quite distant, $z > 4$ (cf. Barger \etal\ 2002). Their survey of the
Chandra Deep-Field, North (equivalent to the HDF(N)) yielded a fair
number of more-distant X-ray identifications; Table 1, below, puts
them in $\Delta z = 0.5$ bins, and includes both narrow and broad-line
(AGN) X-ray sources.

\begin{table}[ht]
\caption{Large Redshifts in the Chandra X-ray Sources (CDFN)}
\begin{center}
\begin{tabular}{cccccc}
\hline
\hline
%\toprule
$\Delta z$ & 2.5-3.0 & 3.0-3.5 & 3.5-4.0 & 4.0-4.5 & $>4.5$ \\
\hline
\bf{n} & 4 & 4 & 1 & 1 & 1 \\
\bottomrule
\end{tabular}
\end{center}
\label{tab1}
\end{table}

A quick inspection of Table 1 and 2, and Fig. 3 suggests no dramatic physical
change in the co-moving density of X-ray emitting galaxies compared to
all field galaxies.

\begin{table}[ht]
\caption{Photometry of Narrow-Line Sources at High Redshift}
\begin{center}
\begin{tabular}{ccccl}
\hline
\hline
%\toprule
Barger (2002) \# & $z$ & R & I & Note \\
\hline
174 & 5.186 & 24.5 & 23.1 & \lya, optically luminous \\
285 & 4.137 & 25.7 & 25.0 & \lya\ emission \\
287 & 2.638 & 24.4 & 23.9 & weak \lya\ \\
294 & 2.240 & 24.1 & 23.5 & weak \lya\ \\
 
\bottomrule
\end{tabular}
\end{center}
\label{tab1}
\end{table}

The largest redshift in the securely-identified group we discuss is
B174 at $z = 5.186$. This source is associated with a moderately faint
optical identification - a bit too faint to classify
morphologically. The near-IR I and z band photometry of this $z \sim
5.2$ source suggest its intrinsic luminosity may lie between that of
luminous QSOs and an ${\mathcal L}^*$ galaxy; the AGN may be partly
hidden, as the spectra of B174 does not display a broad component to
its strong \lya\ emission line. The other three X-ray galaxies at $z >
2$ appear to be residents in normal-luminosity host galaxies, based
upon their photometry. The rough field galaxy correlation between
I mag and the galaxy redshift can be seen in Fig. 3.

\begin{figure}[h!] 
\centering
\epsscale{0.9} 
\vspace{-0.25in}
\plotone{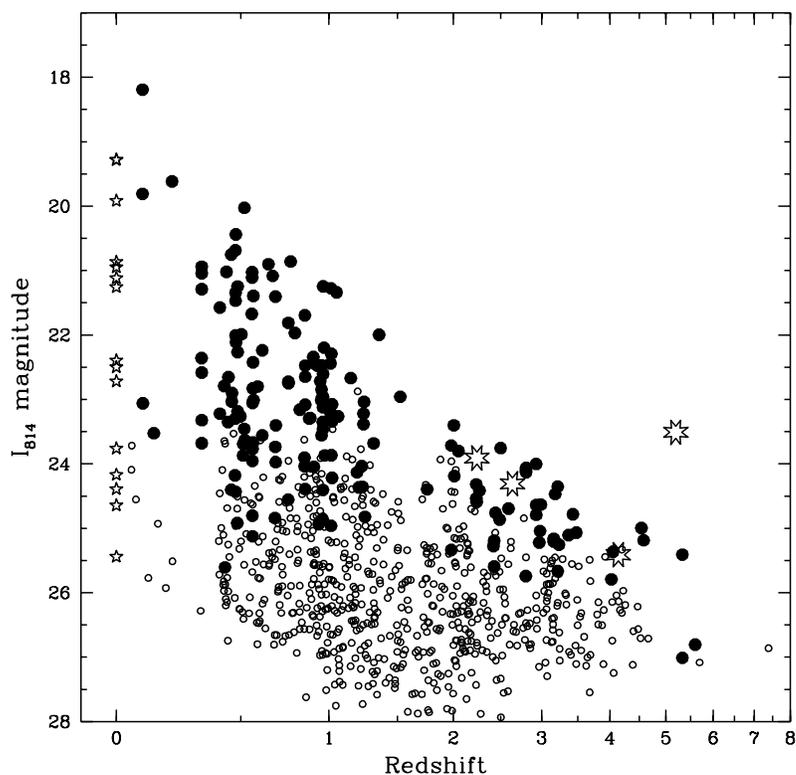}
%\plotone{ewds_sl_int_tidez_lett2.eps}
\vspace{-0.1in}
%\plotfiddle{3b_vs_N.ps}{3.6in}{0}{50}{50}{-150}{-73} % 
\caption{I-band (I814) AB magnitudes versus redshift in the HDF(N),
diagram by Mark Dickinson.  The star-symbols at $z=0$ are Galactic
stars; filled circle symbols have spectroscopic redshifts, while the
small open circles have photometric redshifts.  When spectroscopic
redshifts are found, the photometric point is removed.  We note that
three of the four distant X-ray sources (the large star-symbols) have
rather normal (non-AGN?)  magnitudes - like the field in general.}
\end{figure}

With the present generation of X-ray satellites and plausible
integration times (Mega-secs), we cannot anticipate a large
identification content of X-ray (AGN, or even ``star-burst'') galaxies
beyond $z = 5$. Eventually I would speculate that some sources with
fluxes in the 2-8 keV range below $10^{-16} \ergpscmtw$ may yield a
few very large redshift objects.

\subsection{Dusty Sub-mm (IR) Galaxies}

In recent years it has become evident that a modest number of fairly
high redshift galaxies ($z \ga 2$) are most readily recognized as
unusual at far-IR or sub-mm wavelengths. Our Earth's atmosphere is a
substantial barrier to sub-mm research on galaxies likely to be both
very dusty and also have a rapid pace of star formation. That recipe
augurs for reddened high IR emission, which we can most readily
discover at sub-mm wavelengths. It follows that the most-secure
continuum detections are at a wavelength of 850 $\mu$m (where
receivers are fairly efficient, and our atmosphere fairly
transparent). The detection system of choice at the moment is called
SCUBA, for Submillimeter Common User Bolometer Array. This camera,
utilized on the JCMT (James Clerk Maxwell Telescope), has enabled, for
the first time, deep and relatively unbiased surveys which may
identify the distant dusty galaxies (and/or AGN).

It is important to clarify which galaxies (or AGN) radiate so
profusely at IR and sub-mm wavelengths. They \emph{may} be largely
responsible for the Far-IR extragalactic background. With the
presently available redshifts for securely identified IR/sub-mm
galaxies, their integrated energy density may be quite comparable to
the integrated optical (emitted UV galaxy light measured in the
HDFs) energy (\eg, Genzel \etal 2002).

To deal with this global question, and also to understand the limit of
SFR in a huge star-burst, reddened or not, the crying need is a
reliable set of redshifts.

Blain \etal\ (1999) comment on the reason why many galaxies
detected in the sub-mm spectral window are likely to be at high
redshift. This is because the long wavelength side of the canonical
sub-mm source spectrum has a very steep slope
(cf. Blain \etal\ 2002). The steepness of the long-wave side leads to 
substantial negative K-corrections. That is, the observer's band (850
$\mu$m) benefits from a larger redshift moving the emitted and then
redshifted peak distribution into that atmospheric window. This effect
compensates for the usual geometric dimming of increasing
luminosity-distance at higher z. Thus comparing the sub-mm ($850
~\mu$m) flux with the VLA radio flux (say, near 1.4 GHz) can yield
approximate redshifts without an optical spectrum. But they are not
individually robust.

Another method of deriving a more precise redshift for a sub-mm galaxy
detection is to make good use of the fact that dusty systems
occasionally also show strong molecular lines of CO in emission. The
transitions in CO are (3-2) or (4-3) for the redshift domain of $z \sim
2.6$-2.8 (Frayer \etal\ 1998). But only a small minority have yielded
CO molecular redshifts to date.

Very recently Chapman \etal\ (2003) have succeeded in obtaining good
numbers of optical spectroscopic redshifts for sub-mm galaxies and AGN
with precise radio positions. 16 redshifts of quality were obtained;
probably one is a quasar.  A few others may have some weaker AGN
signal. The median redshift for the galaxies is $z = 2.4$, with a
maximum redshift of $z = 3.699$. Thus one must extrapolate the $850 ~\mu$m
fluxes down to 1-2mJy in anticipation of future achievements in the $z
\ge 5$ domain for sub-mm galaxies. That will surely require new
hardware.

One sort of instrument planned for the near future is the APEX antenna
(the Atacama [Chile] Pathfinder Experiment). It is a planned 12-m
diameter sub-mm telescope at a high, dry site in northern Chile.

Surveys with the APEX should go deeper than the present SCUBA
system. And that will be just a taste of what is to come with ALMA
(the Atacama Large Millimeter Array). ALMA will be the mm/sub-mm
counterpart of the VLT with 64 times the collecting area of APEX! It
should make possible IR galaxy detections 100 times fainter than we
now do with SCUBA and with good spatial acuity. This great array
should lead to many redshifts with molecular CO lines and the
[CII]158 $\mu$m line.

We end this section with an astrophysical speculation: with the
Chapman \etal (2003) data we suggest a relatively high space density
of very luminous and distant sub-mm ($z > 2$) galaxies. They may be 1000
times the density of similarly IR-luminous local star-bursts found
here and now. Hence the detailed study of a few of the powerful IR
galaxies will tell us much about young galaxy SF and dust
interactions.

\subsection{Gamma-Ray-Bursters}

A new and exciting demonstration of extragalactic ``power'' has recently
emerged with the realization that Gamma-Ray-Bursters (GRBs) are
apparently the most powerful cosmic explosions; observing their
optical or radio afterglows can give us an indirect glimpse of a
distant host galaxy. Not all bursters are successfully tracked for
days or weeks after the outburst, but a reasonable fraction do point
to distant $(z \ga 1$) star-burst galaxy hosts. So, for this review we
note that occasional luminous afterglows may signal the locations of
star-forming young galaxies at $z > 4$.

The detailed physics of the situation is unclear, but there are now
believable scenarios suggesting that the GRBs originate from the
collapse of a massive star or even a stellar merger. So sites of
active SF may be one of the ``usual suspects'', much as Type II SNe may
be sited in young-star-rich locations. With the improved ability to
locate GRBs we do find several annual opportunities to follow the
afterglows as they decay; occasionally a redshift from an afterglow
spectrum rich in UV interstellar lines (shifted to the visible) is
obtained. The highest conventional spectroscopic redshift measured to
date is $z = 3.42$ (Kulkarni \etal\ 1998).

Because many GRBs are very luminous (for a short time interval) we
note that the possibility exists to derive ``photo-$z$'s'' or obtain
low-resolution spectra of even more distant GRBs - perhaps with a
little help from their galaxy hosts. Indeed Andersen \etal\ (2000) suggest
a GRB at $z \sim 4.5$ from the afterglow's broad-band colors. At
higher redshifts we will need photometry and/or spectroscopy in the
near-IR. The J-band at $\lambda \sim 1.2$ $\mu$m will take the strong
spectral discontinuity anticipated at \lya\ (1216 \AA, rest) to $z
\sim 9$! Of course our present abilities to obtain good S/N infrared
spectra would be taxed by all but the earliest bright GRB afterglows;
spectroscopy in the first minutes may be needed!

\subsection{Optical Selections of Distant Galaxies: ``Photo-$z$'s'' and \lya\
    Emission Lines}

The case for the use of photometric redshifts -- that is, redshifts
based upon colors in 2 or (likely) more wave bands -- has gradually
strengthened since the mid-1990s. Most critically, we now expect fair
precision from photometric redshifts and few catastrophic failures.

Stern \& Spinrad (1999) compare spectroscopic and photometric
redshifts in the HDF. The photometric redshifts are from the Stony
Brook group (Fernandez-Soto, Lanzetta \& Yahil 1999), and are
determined by fitting the observed galaxy colors (long wavelengths
only for really distant candidates) with redshifted spectral
templates. These templates may be empirical, synthetic, or a hybrid.
A second approach (Connolly \etal\ 1995) is purely empirical - having
already a relationship between previously-observed galaxy redshifts
and the observed total magnitudes (m) with color information (C) to
boot. Then a derived redshift can be found from the multi-dimensional
(m,C) pairs, used for training.  More detail on these ``template
fits'' can be found in the Stern and Spinrad review. Comparisons
between photometric and spectroscopic determination in the HDF yield
residuals typically around 0.1 for $\Delta z$ at almost all redshifts.

Naturally the most important usage of such photometric redshifts is at
very faint levels ($m > 26.5$). These numerous faint galaxies are well
beyond the capabilities of 10-m class telescopes for spectroscopic
redshifts.  The danger here is that galaxies marginally detected in
the red-optical I,z bands and perhaps also in J, H, K [1.2, 1.6,
2.2$\mu$m] can feign very large redshifts if their signal is just a
noise incursion at I or z bands, slightly below the 1 $\mu$m observational
limit of silicon-based CCDs. Since this topic is close to the kernel
of this review, we note that Lanzetta \etal\ (1999) give some examples of
faint, red photometric-z cases of difficult S/N.  Their redshifts
could exceed 6.  Almost all of these ambiguous but potentially
exciting cases have yet to be resolved.  I speculate that better IR
photometry (perhaps using the rejuvenated NICMOS camera on HST) would
help in resolving that situation and perhaps suggest targets for
future generations of near-IR spectrographs.

There is also a systematic problem at some level with color/redshift
degeneracies; blue galaxies in general may show similar colors over a
substantial intermediate z range.  Prior information like the galaxy
apparent magnitude can help decisively.  This ``Bayesian'' procedure
is illustrated by Benitez \& Broadhurst (1999) for the HDF(N).

My personal recent experience with ``I-drops'' (implying a galaxy with
only detectable flux at wavelengths above the I band, $\lambda \ge
8500$ \AA\ at the red edge) is that many of the spectroscopic
candidates (15 to 20 targets per slitmassk) are very difficult due to
their faintness (z$ \sim 25$-26 mag) at longer wavelengths. A few also
turn out to be low-luminosity galactic stars; these late M, L, and T
class dwarfs turn up rather frequently.  Since many of the candidates
come from ground-based imaging, their image structure is not a very
discriminating way to separate stars from QSOs from galaxies.

Most of the I-drops show a marginally detected red-color continuum,
and thus add little to our initial appraisal. It turns out that
approximately a quarter of the I-drops do eventually yield a redshift;
about a third of these with the continuum discontinuity at \lya\
($\lambda_0$ 1216 - the \lya\ ``forest''). Two-thirds of the
spectroscopically detected I-drop systems (with eventual redshifts)
have a noticeable to strong \lya\ emission line. That usually yields
an unambiguous redshift, as the reader can see with the illustrations
in Weymann \etal\ (1998) and Fig. 4, here, by Spinrad, Stern,
Dawson, Filippenko, and the GOODS team ($z = 5.83$).

\begin{figure}[h!] 
\centering
\epsscale{0.9} 

%\vspace{-0.25in}
%\plotone{id3644a.eps}
%\plotone{ewds_sl_int_tidez_lett2.eps}
%\vspace{4in}
\plotfiddle{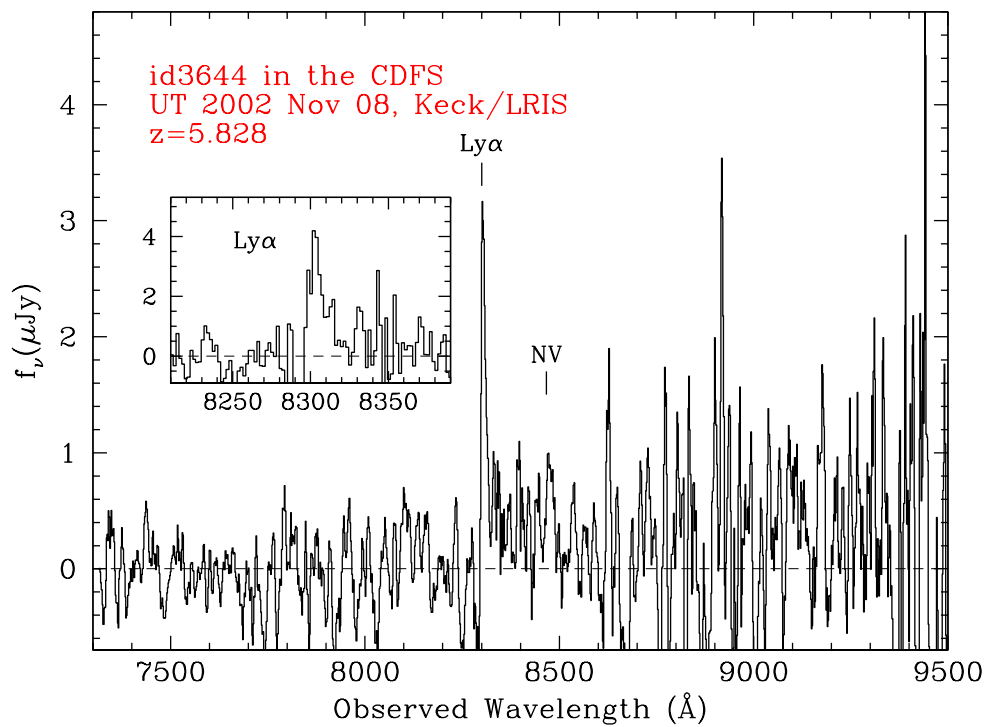}{3.75in}{270}{50}{50}{-210}{300} % 
\caption{ A recent Keck spectrogram of a color-selected (I-drop) faint
galaxy. The strong \lya\ emission line indicates a redshift $z =
5.83$. Also note the continuum discontinuity.  The ``spectral teams''
were led by Spinrad and Filippenko, with reductions by Daniel Stern
and Steve Dawson.  This galaxy was originally selected by Mark
Dickinson and the GOODS team.}
\end{figure}

The pairing of a red continuum color, a continuum discontinuity, and a
fairly strong emission line usually signifies a robust \lya\
redshift. The multiple-criteria spectroscopic technique has been
successful to at least $z = 5.8$ and probably to $z = 6.57$. It should
eventually be pushed to $z \sim 9$ with the \lya\ line at (rest) 12l6
\AA, right in the middle of the conventional near-IR J-band. Right now
that is too technically difficult.

As these pages were being written, two preprints crossed our desk. In
the first, Lehnert \& Bremer (2003) discovered 6 galaxies at $4.8 \le
z \le 5.8$. These galaxies were selected as photometric ``R-drops''-
that is, with little flux in the R-band and a flat spectrum at longer
wavelengths. Follow-up spectroscopy with the VLT yielded accurate
redshifts for these 6, with fairly strong \lya\ emission. Their
largest redshift was $z = 5.869$ (see Table 3 in \S\ref{sec-records}).

The second very timely contribution, by Kodaira \etal\ (2003) (a
Subaru telescope team), used deep narrow-band near-IR images to locate
potentially very distant \lya\ galaxies. The group also obtained a few
spectra which lead to two fairly certain identifications.  One line,
with a symmetric line shape, is assumed to be \lya\, and in the other
case it appears to be satisfactorily asymmetric, hence reliably
\lya\ (see \S\ref{sec-lya} for discussion of this point). The best
spectrum is of SDFJ 132418.3 at $z = 6.578$. That would make this
\lya\ galaxy the largest redshift of any individual system measured to
date. The redshift is only slightly greater than that of HCM 6 A ($z =
6.56$) by Hu \etal (2002), and Hu, Cowie \& McMahon (2002).
\begin{figure}[h!] 
\centering
\epsscale{0.9} 
\vspace{-0.25in}
\plotfiddle{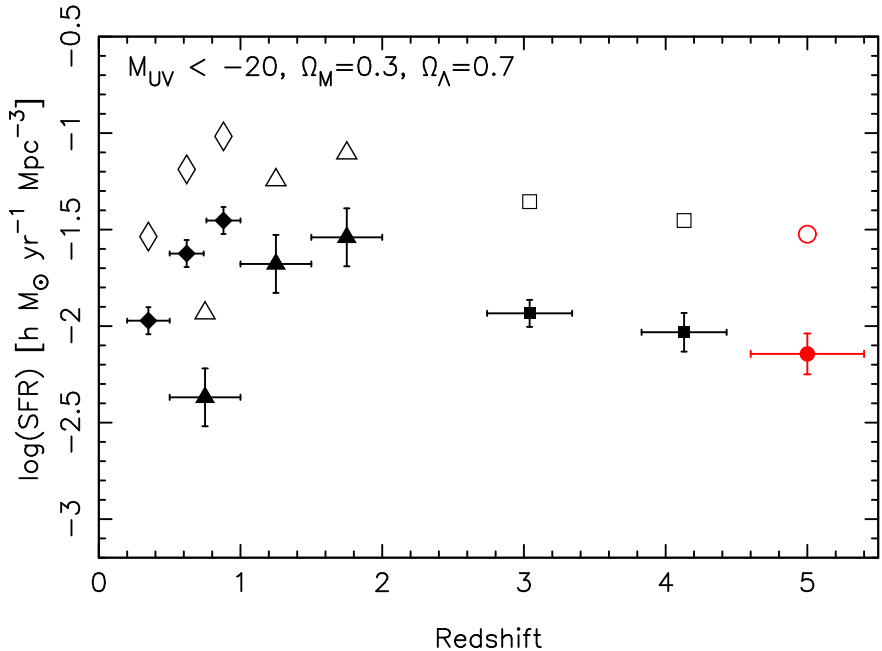}{3.6in}{270}{50}{50}{-184}{290} % 
%\plotone{madau_plot.eps}
%\plotone{ewds_sl_int_tidez_lett2.eps}
%\vspace{-0.1in}
%\plotfiddle{3b_vs_N.ps}{3.6in}{0}{50}{50}{-150}{-73} % 
\caption{Star-formation rate density as a function of redshift based
on the UV-luminosity function with a magnitude limit $M_{UV} < -20$.
Triangles and diamonds are from Connolly \etal\ (1997) and Lilly \etal
(1996), respectively.  Squares represent data from Steidel \etal\
(1999) at $\langle z \rangle \sim 3$ and and 4.  The circle is the
data of Iwata \etal\ (2003) at $z=5$.  Filled symbols indicate values
without correction for dust extinction.  Dust extinction was corrected
following the prescription of Calzetti \etal\ (2000) with ${\rm
E(B-V)}=0.15$ for all data points.  Dust-corrected values are denoted
by open symbols.  Plot courtesy of I. Iwata. }
\end{figure}

These very contemporary detections of galaxies beyond the
``QSO-limit'' of $z = 6.4$ show us that UV emission from galaxies is
still present at the ``tail'' of the ``dark ages''.  A future
space-desideratum will be the galaxy morphology \emph{in} the \lya\
line.  We are interested in any extended neutral gas about the galaxy
-- via the scattered \lya\ emission from the central ionizing
region (Haiman 2002 and references therein).

When the luminosity function of \lya\ emitters is extended to $z \sim
6.5$ (fainter galaxies have to be included) we should be able to
extend the SFR density to that great distance.  A sample of the
near-constancy of the SFR density from $z \sim 2$ to $z \sim 5$ is
illustrated in Fig 5 (from Iwata \etal\ 2003).  The galaxies going
into the computation of the SFR density are photometrically selected,
using the top of the UV-luminosity function ($M_{UV} - 5 \log{\h} <
-20$).  Interestingly, an attempt by D. Stern and the author to
utilize serendipitously discovered \lya\ emitters at $z \sim 5$ yields
a SFR density slightly higher than that of the $z\sim 5$ Iwata point
in Fig. 5 (with considerable uncertainty).  We view this as a possible
coincidence, as these two methodologies may be sampling different
populations.  It is somewhat surprising that the relatively slight
decline of the SFR density, noted by Iwata \etal (2003) should be
maintained to $z \sim 5$.  At that redshift the detected objects are
effectively sub-galactic in size and probably rather modest in
mass. At least temporarily, their M/L ratios must be quite low.  Will
that be true of most small sub-galactic systems?

\subsection{Details on the \lya\ Emission Line in Very Distant Galaxies} \label{sec-lya}

The classic proposal by Partridge \& Peebles (1967) that the \lya\
emission line might carry a fair fraction of the escaping bolometric
luminosity of a young-star-rich galaxy is now testable.  The 
review by Pritchet (1994) is also strongly recommended.  Of course
these early predictions did not reflect the possible presence of dust.
Since the 1990s various searches have been initiated for \lya-emitting
galaxies at large redshifts.  Initially all of these searches led to
negative results (eg, Thompson \& Djorgovski 1995).

However, deeper photometric and spectroscopic searches of the last 6-7
years have yielded a modest number of ``safe'' \lya\ emitters - often (at
the largest $z$'s) the line being the only measurable spectral
feature. The peak flux from a distant \lya\ emission line galaxy can
often exceed the (redward) continuum level by a factor greater than
10! Of course the line from a faint system still has to compete with
the strong telluric sky emission bands of OH and O$_2$. Space-spectra
won't deal with such a bright near-IR sky, and that will be advantageous.

Successful \lya\ searches include Cowie \etal\ (1998); Hu \etal\ (1998),
Pascarelle \etal (1998), Hu \etal\ (1999), Steidel \etal (2000),
Kudritski \etal (2000), Fynbo, M\"oller, and Thomsen, (2001). 

There are three modes of \lya\ detection used with success in the past
few years. They are narrow-band photometric excesses at fixed
wavelengths (redshifts), a \lya\ forest (Lyman breaks in the continua)
plus emission at the line, and serendipitous or fortuitous detections
on multi-slit spectrograms. The issues we may face for each/all of the
sub-types include the emission line strength and shape, the luminosity
function of \lya\ emitters (and their surface densities), the effect
of widespread neutral gas and dust, and the termination of the ``dark
ages'' before or during the re-ionization epoch. Many of these topics
have been addressed recently by Stern \& Spinrad (1999); Rhoads \etal\
(2003); Ellis \etal\ (2001); Hu \etal\ (1999); Hu \etal\ (2002a), and
in a predictive manner by Stiavelli (2002).

I suggest a few specific points where new observations and
interpretations may be of substantial interest. For example, we'd like
to confirm or deny that strong emission line \lya\ galaxies ($z \ge
4$) obey the same luminosity function distribution as do
photometrically selected Lyman break systems at $z = 3$ and $z = 4$
(\cf. Steidel \etal\ 1999; Giavalisco 2002).

The difficulty in a present-sample comparison between Lyman break
galaxies and \lya\ emitters is that (at high luminosities, at least)
only a modest fraction of Lyman break (continuum selected) galaxies
have strong \lya\ emission lines ($W_0 > 20$ \AA, say). Among the
\lya-emitting systems (narrow-band or serendipitous detections) many
candidates have very faint continua and would be missed in normal
broad-band photometry. This latter bias is stressed by Fynbo \etal\
(2001). Indeed, Rhoads \etal\ (2003) found that if they summarized the
line/continuum ratio in \lya\ galaxies, the equivalent widths
occasionally ``rose'' to $W_{\lambda}^0 \ge1000$ \AA\, but more
frequently to 190 \AA.  60\% of the \lya\ emitters studied by Malhotra
\& Rhoads (2002) had observed equivalent widths (hereafter EW) $ >
240$ \AA. For Ly-break systems, Shapley \etal\ (2001) find their
60th-percentile line to be a marginally-detectable 20 \AA\ EW.  The
Shapley galaxies are at a slightly lower redshift; that difference is
not critical.

If the above trend of lower-continuum-luminosity galaxies ($z > 4$)
having stronger \lya-emission were to continue, we might diagnose this
systematic as a trend toward lower metallicities for lower masses. But
there are other possibilities; the \lya-emission line may as easily
depend upon physical outflows (galactic winds), which in turn could
have some total mass-dependence (or merger timing).

 To get some idea as to the evolution of the luminosity function of
young galaxies, we can compare the surface densities of distant
galaxies.  Pritchet (1994) made a first approximation to this. We
utilize the Steidel \etal\ (1999) luminosity function zero point, and the
``predictions'' by Lanzetta \etal\ (1999) and Stern \& Spinrad (1999) for a
constant (with $z$) luminosity function. The cumulative surface
density of identified $z \ge 4.5$ galaxies in the HDF(N) is about
$1.5/\Box^{\prime}$.  These galaxies constitute a sample of continuum
galaxies (photo-zs) and emission line galaxies with $I_{814} \le
26.5$. This is very close to the ``prediction'' of the Lanzetta
(unevolved) surface density (also see Ouchi \etal\ 2002).

The Lanzetta (1999) surface density curves do suggest a drop in
the faint galaxy surface densities for the extreme case, $z \ge 6$;
that is not surprising at about $I_{814} = 26$.  Still at slightly fainter
magnitude levels a measure of the $z \ge 6.0$ density by
broad-band/narrow-band photometry may be a viable check on the
luminosity function zero point and its shape (Lehnert \& Bremer 2003).

What is the best physical interpretation of the very large EWs of
\lya\ often measured for galaxies at $z > 3$?

The \lya-emitting galaxies with line EW in excess of 200 \AA\
(rest-frame) (Malhotra \& Rhoads 2002) are difficult to explain with a
conventional O-B star mass function and ionizing spectra that are
similar to those anticipated in extant solar-abundance models. The
models rarely (and temporally) exhibit $W_{\lambda}^0 \ge 150$ \AA
(\eg, Charlot \& Fall 1993). To decrease the observed \lya\ EW would be
easy; as the dominant resonance line it is scattered frequently, and
the resulting ``random spatial walk'' at the center of this line,
coupled by small amounts of dust, can easily and drastically reduce
the emission measure. It would, of course, also depend on the
geometry.

To obtain a higher EW and/or higher flux in \lya, one can call upon
three scenarios:

(a) A ``tilted'' mass function, with more O stars than found in local
HII regions, as an \emph{ad hoc} premise.

(b) We can also reduce the heavy element abundances in our models, and
this allows an increase in the number of ionizing photons per O
star. A recent paper by Schaerer (2003) considers the temporal
evolution of the \lya\ line from model stellar populations ranging
down from solar metal-abundances to very low metallicities (below the
abundance level of the most metal-poor stars and gas in relatively
nearby star-forming systems). We amplify this discussion below.

(c) Finally, sometimes a strong \lya\ emission line is the signature
of an AGN.  However, ``real'' AGN spectra, from QSOs down to
modest-luminosity accretions, usually produce a broader \lya\ emission
line ($\Delta v \ge 1000$ \kms) than seen in normal galaxies ($\Delta
v \sim 500$ \kms). They usually, but not always, also show \civ\
(moderately broad) 1549 \AA.  So most of the narrow-line \lya\
galaxies must have a line powered by the UV flux from OB stars.  This
is confirmed by the lack of hard X-ray flux in LALA galaxies at $z
\simeq 4.5$ (Malhotra \etal\ 2003), indicating they are not obscured
AGN.

The previously-mentioned Schaerer paper (Schaerer 2003) predicts EW
of $\sim$ 240-350 \AA\ for metallicities down to $Z = 4 \times
10^{-4}$ (down from solar by a factor of $\sim 50$ times).
Stiavelli (2002) shows even larger EW for \lya\ in metal-poor OB
stars.  Conceivably the initial stellar mass function (IMF) could also
vary and be itself slanted toward higher masses because of the lower
abundances. So the pairing of low abundance and a structure favoring
massive O stars might allow EW to match most of the \lya\ galaxies
selected by Malhotra \& Rhoads (2002) and by Rhoads \etal\ (2003). An
almost-practical spectroscopic test of this idea can be made by
examining the UV HeII transition at $\lambda_0$1640 \AA. This line is
much weaker than \lya\ in star-forming populations - with EW $\sim5$
\AA\ anticipated at low abundances of the metals. At higher abundances
(near solar) it will be even weaker. Thus higher S/N spectrograms will
be required in practice to use this \hetwo\ feature in \lya\ ``test
galaxies''.

The shape of the \lya\ emission line in distant star-forming galaxies
is peculiar and may turn out to be an interesting guide to the
circumgalactic medium as well as to galaxian winds or sporadic
outflows.

The asymmetry of the \lya\ line has been noted by Kunth \etal\ (1998)
and Pettini \etal (2001); it is also mentioned by Stern \& Spinrad
(1999). We have utilized the broad red wing of the \lya\ line and its
sharp ISM/IGM cutoff on the blue side as a secondary criterion for
assuming a single strong emission line is to be identified as
\lya. This is opposed to the profile of the [\otwo] $]3727$ doublet --
unresolved in most lower-spectral-purity observations of faint
objects. Recent work by E. Landes, S. Dawson, and the author has
compared a spectal asymmetry index (a lambda-space ratio) for ten
strong \lya\ emission lines; this particular index is small for a
symmetric line and large for a red winged emission.  Out of a sample
of seven medium-resolution spectra of galaxies with a ``solid''
[\otwo] identification ($z = 1.0$) the asymmetry index averages $0.9
\pm 0.1$, while the 10 bonafide \lya\ galaxies, with $\langle z
\rangle \approx 4$ display a larger range of index, from 1.0 to 2.3,
with none less than unity.  Seven of the \lya\ systems are clearly
asymmetric with a noticeable red wing (see Fig. 6).
 
\begin{figure}[h!] 
\centering
\epsscale{1} 
%\vspace{-0.25in}
%\plotone{0lam_30_r1_separate.ps}
%\plotone{ewds_sl_int_tidez_lett2.eps}
\plotfiddle{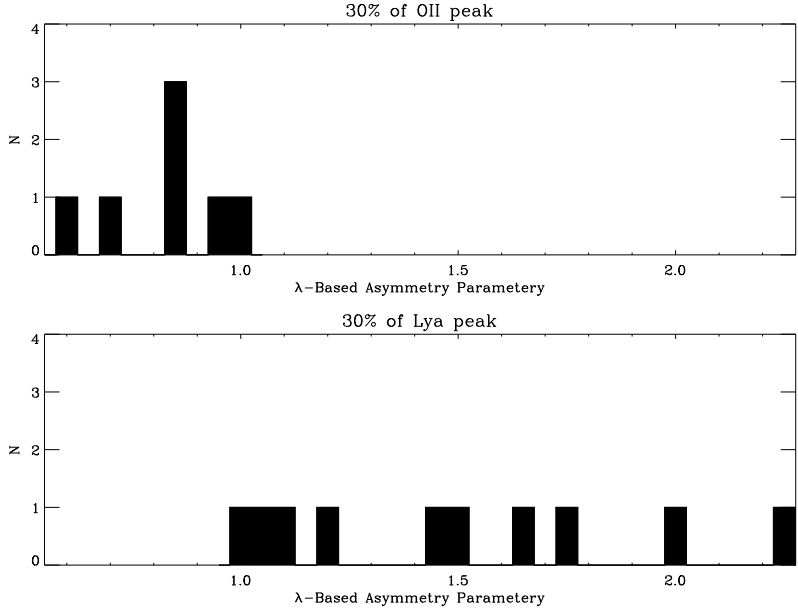}{3.6in}{90}{50}{50}{188}{-25} % 
%\vspace{-0.1in}
\caption{The \lya\ emission line asymmetry index, applied to [\otwo]
emitters (upper panel) and to \lya\ lines (lower panel). Ther line
wavelength asymmetry is defined at 30\% of the line peak; an index
over unity implies a stronger red wing to the line profile.  Most but
not all of the strong \lya\ line emitters show an asymmetric red wing,
with an index $\ge 1.5$.  The \lya\ galaxies range in redshift from
$z=3$ to $z=5.3$.  Figure and reductions by Emily Landes.}
\end{figure}

The \lya\ line is usually steeply declining on its blue side; we'll
soon come back to this observation. So deciding whether an emission
line is [\otwo] at a modest $z$ or \lya\ at a large $z$, can often be
helped by measuring the asymmetry. Of course a \lya\ (bigger redshift)
decision based upon a large line asymmetry index becomes a sufficient,
but not necessary condition for claiming the \lya\ identification.

The astrophysics behind the red wing of \lya\ has been well expounded
by Tenorio-Tagle \etal\ (1999), Ahn, Lee \& Lee (2002), and Dawson
\etal (2002). The scenario here is a mini-galaxy scale outflow of
neutral and partly ionized matter; the blueward velocity component
being absorbed by external and expanding neutral H gas between us and
the outflow.  The backscattered component can be sufficiently
redshifted off of the receeding wind, and hence avoid immediate
absorption.  This will impose a broadened red wing to the \lya\ line.

On the blue side of \lya\ we have a rapid decrease in intensity, a very
sharp cutoff to the galaxy emission line at a slightly smaller
redshift. The actual galaxy systemic velocity is likely to be near but
blueward of the line peak, rather than its bisector at about half of
maximum intensity.

In any case the \lya\ H absorption can take place in neutral
circumgalactic gas, and in putative cluster gas, and also, at slightly
lower redshift, neutral H clouds in the IGM - the well-studied \lya\
forest.

One interesting semi-quantitative aspect of the blue side cutoff is
the difference we have noticed between the blue edge of \lya\ in QSO
spectra and that of the normal distant galaxies, highlighted in this
review (see Fig. 7).  A new type of ``proximity effect'' seems in
place, in the sense that the galaxy \lya\ profile on the short
wavelength side is extremely steep, going from the line peak to near
zero intensity in $\Delta v_1 = 100$ \kms, on our few echelle (higher
spectral resolution) observations of the brightest distant systems (in
their \lya\ line). The profile on the blue side of the strong emission
line in QSO spectra (also $z > 4$) is moderately steep, but has a
typical $\Delta v_2 \approx 800$ \kms, but often $>1000$ \kms.

\begin{figure}[h!] 
%\centering
\epsscale{1} 
%\vspace{0.25in}
\plotone{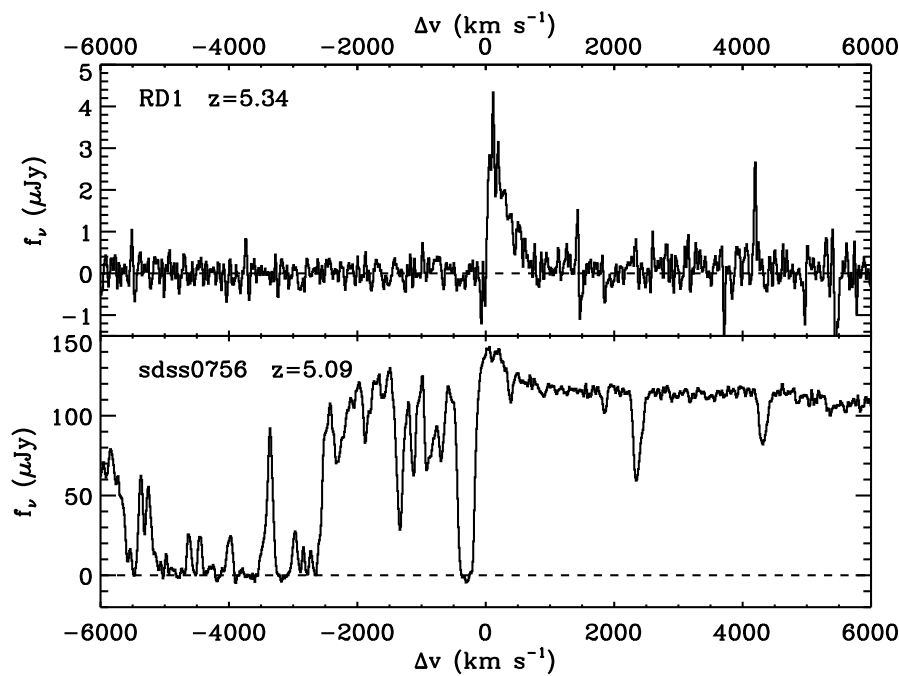}
%\plotone{ewds_sl_int_tidez_lett2.eps}
%\vspace{-0.1in}
%\plotfiddle{
\caption{The steepness of the ultraviolet side of the \lya\ emission
line in a QSO ($z=5.09$) and a faint galaxy, RD1 ($z=5.34$,
Dey \etal\ 1998).  The very sharp and rapid decline of the blue side in
the distant galaxy may be indicative of nearby (surrounding?) neutral
gas.  On the other hand, the QSO presumably ionizes much of any
circumgalactic H originally present (with a small $\Delta v$) Thus the
QSO line and continua are detectable to $\Delta v_2 \simeq 2500$ \kms.
Reductions and Figure by S. Dawson.}
\vspace{-.25cm}
\end{figure}

Our interpretation of this systematic difference between UV-luminous
QSOs and UV-fainter galaxies is straightforward. In proximity to the
luminous ultraviolet radiation field of the QSOs H is very thoroughly
ionized and thus doesn't absorb \lya\ photons at small $\Delta v$. On
the other hand, a galaxy's UV ionizing radiation may not escape (or
fully escape - see Dawson \etal\ 2002). Thus the rapid decline on the
blue side of \lya\ may simply augur the existence of neutral gas in
the circumgalactic environment near the galaxy. The effect may
increase with redshift, but this is not yet well documented. This
trend is potentially of interest in our present and future attempts to
document the degree of IGM ionization near active objects and also on
a diffuse, larger scale. Our coverage in redshift implies that we are
looking back close to the re-ionization redshift, between $z = 6$ and
$z = 20$, apparently.

\subsection{ Current Redshift Record Breakers With \lya\ Emission or Absorption Breaks} \label{sec-records}

In Table 3 we list published or otherwise secure ``record redshifts''
for galaxies; most have prominent \lya\ emission lines or at least a
strong \lya\ forest absorption.  \small

\begin{table}[h!]
\caption{Census of Galaxies Confirmed at $z \ga 5$}
\begin{center}
\begin{tabular}{lllccccc}
\hline
\hline
%\toprule
z & Source & Reference & NB & LBG & ser & other & lens \\
\hline
%\hline
6.578 &  SDFJ 132418.3 &  Kodaira et al. & x &  & & & \\
6.56 & HCM 6A &  Hu etal. & x & & & & x \\
6.541 & SDFJ 132415.7 & Kodaira et al. & x & & & & \\
5.869 & BDF1:19 & Lehnert and Bremer & & x & & &  \\
5.83 & CDFS 5144 & GOODS & & x & & & \\
5.783 & CDFS SBM03\#3 & Bunker et al. 2003 & & x & & & \\
5.746 & LALA5 1-03 & Rhoads et al. & x & & & & \\
5.744 & BDF1:10 & Lehnert and Bremer &  & x & & & \\
5.74 & SSA22-HCMI & Hu etal. & x & & & & x \\
5.700 & LALA5 1-06 & Rhoads et al. &  x & & & & \\
5.69 & LAE J1044-0130 & Ajiki et al. 2002 &  x & & & & \\
5.674 & LALA5 1-5 &  Rhoads et al. 2002 & x & & & & \\
5.655 & LAE J1044-0123 & Taniguchi et al. & x & & & & \\
5.649 & BDF2:19 &  Lehnert and Bremer & & x & & & \\
5.631 & HDFF 36246-1511 &  Dawson et al. 2001 & & & x & & \\ 
5.621 & Lynx R-drop & Stern et al. in prep & & x & & & x \\
5.60 & HDF 4-473 &  Weymann et al. 1998 &  & x & & & \\
5.576 & Abell 2218 lens & Ellis et al. 2001 & & & x & & x \\
5.46 & NDFWS R-drop & Dey et al., in prep. & & x & & & \\
5.34 & HDF 3-951.0 &  Spinrad et al. 1998 &   &  x & & & \\
5.34 & RD1 & Dey et al. 1998 & & & x & & \\
%5.221 & Pisces R-drop & Stern et al. in prep  & & x & & & \\
5.190 & HDFF ES1 & Dawson et al. 2001 & & & x & & \\
5.19 & TN J0924-2201 &  van Breugel et al. 1999 & & & & x & \\
5.19 & ES1 & Dawson et al. 2001 & & & x & & \\
5.186 & HDFF Chandra source & Barger et al. 2002 & & & & x & \\
5.12 & A1689 lens & Frye et al.   2002 &    & x  &   &   & x \\
5.056 & BDF1:26 & Lehnert et al. 2002  &   & x  &   &   &    \\
5.018 & BDF1:18 & Lehnert et al. 2002  &   & x &   &   &    \\
4.99 & Cetus R-drop &  Stern et al. in prep & & x & & & \\
\bottomrule
\end{tabular}
\end{center}
{Notes on the initial discovery techniques -- NB = narrow-band
selected; LBG = continuum Lyman-break/Lyman-forest break selected; ser
= serendipitously identified; other = selected in other manner (e.g.,
radio-selected, X-ray selected); lens = known gravitational lens.}
\label{tab1}
\end{table}
\normalsize

We note that since 1999 astronomers have added at least 25 galaxies
with $z \ge 5$. This is an impressive and useful score; however, a
more physical analysis of several aspects of the pioneering effort is
now an obvious and desired second approach. Also, the morphologies of
the continuum and \lya\ lines may provide useful information on the
environs of very early galactic systems.  The cut-off date for entries
in Table 8.3 was 2003 February.

\section{The Future} \label{sec-future}

Wide-field narrow-band and broad-band imaging with large ground-based
telescopes have considerable promise. Narrow-band and
broad-spectral-band studies of the sky areas already earmarked for
multi-wave observation is one useful approach. It is already been
successful in the Hubble Deep Fields. Such imaging photometry has
already turned up distant galaxies, especially at $z = 5.7$ and $z =
6.6$ (airglow windows for narrow-band studies). We know of several
groups planning to search for \lya\ emitters at the highest redshifts
available to CCD detectors ($\lambda \approx 9200$ \AA; $z_{\alpha} =
6.6$). A more ambitious plan would be to utilize IR detectors at the
best (OH-band-free) sky windows in J band ($\lambda \approx 12,000$
\AA; $z = 9$). Exploration of the interval $6.6 \le z \le 9$ should
bring us to the edge of the re-ionization epoch where the first stars
and quasars began to ionize (again) the halos around collections of
dark matter and baryons. A schematic cartoon (Pentericci \etal\ 2002b)
is shown by Loeb \& Barkana (2001). Is it realistic? We'll hope that
very distant galaxy images and spectra will tell us about very early
star formation at the end of the long ``dark age''. At this time it is
uncertain as to whether the first luminous and ionizing objects were
star-forming galaxies! But something or some process began
star-formation through the darkness and led to the formation of young
stars and young galaxies. We may soon barely detect these faint
``first galaxies'' with our telescopes and intellects.

\section{Acknowledgments}
I thank Curt Manning, Steve Dawson, Arjun Dey, Mark Dickinson, Ikuru
Iwata, Emily Landes, Scott Chapman, and Dan Stern for help
with the science, and with this manuscript .  I
acknowledge the support from NSF Grant AST--0097163.
%\newpage

%\bibliographystyle{apj} 
%\bibliography{apj-jour,/tortilla/manning/hy/manning}

%\end{document}

\end{document}